\documentclass[prl,two column, showpacs]{revtex4}
\usepackage[T1]{fontenc}
\usepackage[latin1]{inputenc}
\usepackage{graphicx}
\usepackage{amsmath}
\usepackage{amssymb}
\def\g{\gamma}
\def\e{\varepsilon}
\def\d{\delta}

\def\te{\theta}

\def\s{\sigma}

\def\D{\Delta}

\def\bk{{\bf k}}

\def\bm{{\bf m}}

\def\be{\begin{equation}}
\def\ee{\end{equation}}
\def\bea{\begin{eqnarray}}
\def\eea{\end{eqnarray}}
\def\nn{\nonumber}
\def\lb{\label}

\begin{document}

\title{Quantum effects for ballistic transport in spintronic devices}
\author{H.G. Silva$^{1,2}$, Y.G. Pogorelov$^1$}
\affiliation{$^1$IFIMUP, Universidade do Porto, R. Campo Alegre, 687, Porto 4169-007,
Portugal,\\ $^2$CEOT, Universidade do Algarve, Campus de Gambelas, Faro 8005-139, Portugal}

\begin{abstract}
Recent fabrication of atomic precision nanodevices for spintronics greatly boosted their
performance and also revealed new interesting features, as oscillating magnetoresistance
with number of atomic layers in a multilayered structure. This motivates the need to go
beyond the usual theoretical approach of semi-classical continuous layers. Here the simple
tight-binding dynamics is used to describe quantum conduction in a multicomponent system
with spin-polarized electrodes separated by an ultrathin and atomically coherent non-magnetic
spacer (either metallic or insulating). A possibility is indicated for obtaining a huge
resonant enhancement of magnetoresistance in such device by a special choice of gate voltage
on the spacer element.
\end{abstract}


\maketitle\

\section{Introduction}

In our information based society the development of ultra-high density storage technology
is a demanding priority. In this context, the necessity in ultra-high sensitivity read-head
devices is a great challenge from both theoretical and experimental points of view. Presently
the most promising candidates for this purpose are the magnetic tunnel junctions (MTJ) made
by two magnetic electrodes separated by an ultra-thin non-magnetic spacer and their study
becomes one of the central topics in the fast developing field of spintronics. Since the
early studies by Tedrow and Meservey \cite{tedrow1} on spin polarized tunnel conduction, an
impressive progress was achieved either in experiment \cite{stearns, meservey, moodera} and
in theory \cite{buttiker, schep, mathon1} for the spintronics applications of this mechanism.
The most important recent advances are related to nano-fabrication of multilayered systems on
atomic precision level \cite{yuasa1, parkin}, which rises the MTJ performance up to ~400$\%$
of magnetoresistance and enables a breakthrough to their fundamental quantum properties. It
should be noted that the overall number of electronic degrees of freedom in a device like
MTJ is macroscopically big which generally suggests a quasi-classical behavior, controlled by
the spin-dependent relaxation times or by the spin-dependent tunneling amplitudes. But the
essentially quantum behavior turns to be possible at effective separation of a small number
(few units) of electronic degrees of freedom among the macroscopically big total number as,
e.g., the hoppings between the planes in the spacer among all possible hoppings in a junction,
forming a partial quantization of energy spectrum and drastically enhancing the sensitivity
of tunnel (or ballistic) transport to external factors \cite{sun}.  Another natural
quantization effect is the oscillatory behavior of conductance, either in function of the
spacer thickness (or, more exactly, the number of atomic planes) and in electric field on the
junction \cite{yuasa2}, which may also allow an interesting possibility for specific
magnetoconductance oscillations. All this needs that the mode mixing due to the interface
roughness and intra-spacer defects be below the characteristic energy quantization scale,
and practically requires that the spacer consist of few atomic planes, coherent enough.

Consequently, the theoretical analysis of such systems requires a fully quantum-mechanical
description, rather than more traditional semi-classical approaches \cite{sim,julliere}.
Up to the moment, there already exists a rather well elaborated theoretical base for this
description, using the Green function formalism and rigorous \emph{ab-initio} band calculations
\cite{butler, mathon, mathon2} as inputs to the general Kubo's formula. However, in many
practical cases the direct use of corresponding algorithms leads to heavy enough numerical
work, specific for each particular configuration and not very well suited for qualitative
predictions and optimization of device performance.

In this paper instead, the simple tight-binding dynamics in single-band approximation is
developed, using the straightforward equations of motion for on-site quantum-mechanical
amplitudes, to get a handy description of quantum magnetotransport in the ballistic regime
(absence of either thermal or impurity scattering) for a trilayer system of spin-polarized
electrodes with an ultra-thin and atomically coherent non-magnetic spacer. The motivation
for our approach is an easy generalization to more promising device geometries (double barriers
or double junctions, etc.) and conduction regimes (including finite electric field effect)
which will be presented in a forthcoming work. This presentation is mainly limited to the
basics of the method and to its most characteristic results. Thus in Sec. 1 the explicit quantum
wave functions are obtained for the $1$-dimensional isolated atomic chain. In the following
Sec. 2 the finite $1$D chain is inserted between two $1$D semi-infinite leads and the
transmission and reflection coefficients for a collective electronic state are analytically
calculated. Further, in Sec. 3 this result is generalized to the $3$-dimensional case and the
Landauer conductance formula \cite{landauer} is used in the $3$D version to yield a clear
picture of basic quantum effects evolved in this coherent system. In Sec. 4, the important
effects of electronic correlation are included into consideration using the approximation of phenomenological interface potentials, which foresees a more consistent treatment in the spirit
of density functional theory. At last, in Sec. 5 a work summary and the principal results are
presented and commented.

\section{1. Basic chain model}

\lb{bas}The simplest model for transport over exact electronic states considers a linear chain of
$n$ identical atoms with single available electronic state $|l\rangle$ on each $l$-th atomic
site and describes the single-electron dynamics in the simplest tight-binding approximation
with (real) hopping amplitude $t$ between nearest neighbor sites (taking the distance between
them as unit length).

\begin{figure}
\center\includegraphics[bb=400bp 250bp 450bp 430bp,scale=0.4]{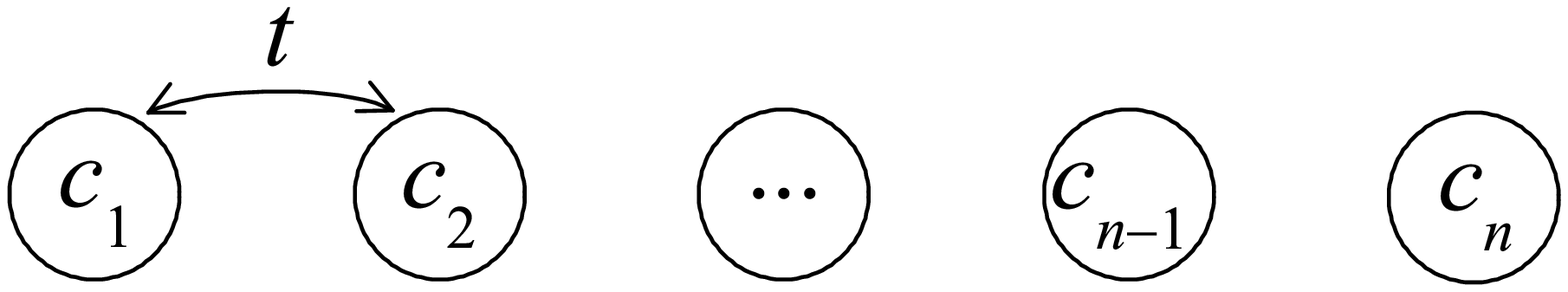}
\caption{\lb{cap:1}Finite atomic chain with tight-binding amplitude $t$.}
\end{figure}

In this coupled chain, any collective electronic state can be expressed as $|c\rangle =
\sum_{l=1}^n c_l |l\rangle$, with complex amplitudes $c_l$ and atomic states $|l\rangle =
\hat c_l^\dagger | 0 \rangle$, generated by the second quantization operators acting on the
vacuum state $| 0 \rangle$. Choosing the on-site atomic energy as a reference ($\e_c = 0$),
we write the Hamiltonian operator as:

\be
  \hat H^{(n)} = t \sum_{l=1}^{n-1}(\hat c_l^\dagger \hat c_{l+1} + \hat c_{l+1}^\dagger
   \hat c_l)
    \lb{eq1}
     \ee

\noindent and obtain the electronic spectrum $\e_m$ ($m = 1,\dots,n$) as the roots of the
secular equation $D_n(\e) = \det(\e - \hat H^{(n)})= 0$ with the corresponding Hamiltonian
matrix $H_{l,l^\prime}^{(n)} =\langle l|\hat H^{(n)} |l^\prime\rangle = t \left(\d_{l,l+1}
\te_{l-1} + \d_{l,l-1}\te_{n-l}\right)$ (where $\d_{l,l^\prime}$ is the Kronecker delta and
$\te_l = 1$ if $l > 0$, otherwise zero). These determinants satisfy the recurrent relation:

\be
 D_n(\e) = \e D_{n-1}(\e) - t^2 D_{n-2}(\e),\quad n\geq 2,
   \lb{eq2}
     \ee

\noindent with the initial conditions $D_0(\e) = 1,\,D_1(\e) = \e$, that define them exactly
through the $2$nd kind Chebyshev polynomials: $D_n(\e) = t^n u_n(\e/2t)$ \cite{abramowitz}.
Then it is convenient to pass to these dimensionless polynomials $u_n(x)$ as functions of
the dimensionless variable $x = \e/2t$, rewriting Eq. \ref{eq2} as:

\be
 2x u_n(x) = u_{n+1}(x) + u_{n-1}(x),
  \lb{eq3}
    \ee

\noindent with $u_0(x) = 1,\, u_1(x) = 2x$. A useful trigonometric parametrization $u_l(\cos \te)=\sin[(l+1)\te]/\sin \te$ permits to present the general solution of Eq. \ref{eq3} as:

\be
 u_l(x) = \frac{\sin \left[(l+1)q_x\right]}{\sin q_x},
  \lb{eq4}
    \ee

\noindent where $q_x = \arccos x$. Then the discrete energy spectrum resulting from zeros of
$u_n(x)$ is explicitly given by:

\be
 \e_m = 2 t \cos \frac{\pi m}{n+1},\quad m = 1,\dots,n.
   \lb{eq5}
     \ee

\noindent Now let $c(x) = \left(c_1(x), \dots, c_n(x)\right)$ be the eigen-vector of the
Hamiltonian matrix, Eq. \ref{eq2}, related to the eigen-energy $\e = 2tx$ (in what follows we
mostly drop the explicit energy arguments of amplitudes like $c_l$). Its components satisfy
the tight-binding equations of motion

\be
  2x c_l = c_{l+1} + c_{l-1},\quad{\rm for}\quad 2 \leq l \leq n-1,
   \lb{eq6}
    \ee

\noindent completed by $2x c_1 =  c_2$ and $2x c_n = c_{n-1}$. Since Eq. \ref{eq6} for
$c_l/c_1$ is just equivalent to Eq. \ref{eq3} for $u_{l-1}$, the eigen-vector components can
be written as:

\be
  c_l = \frac {\sin\left(l q_x\right)}{\sin q_x}c_1.
   \lb{eq7}
    \ee

\noindent We notice that this solution also satisfies the above mentioned equations of motion
for $c_1$ and $c_n$ and provides the "closed" boundary conditions:

\be
 c_0 = c_{n+1} = 0.
  \lb{eq8}
   \ee

\noindent As usual, the value of $c_1$ is fixed by the normalization condition, $\sum_l
|c_l(x)|^2 = 1$, giving finally the $l$-th component of the eigen-vector (related to the
eigen-energy $\e_m = 2tx_m$) as:

\[c_l\left(x_m\right) = \sqrt{\frac 2 {n+1}} \sin\frac{\pi m l}{n+1}.\]

\noindent Our next purpose is to consider this finite chain inserted into the "circuit"
between two semi-infinite chain leads.

\section{2. Transmission through discrete chain structure}

\lb{chain} For a composite system of finite $n$-chain (in what follows called the gate, G) between
two semi-infinite chain leads, S (source) and D (drain) (Fig. \ref{cap:2}), the tight-binding
Hamiltonian, Eq. \ref{eq1}, is extended to: $\hat H = \hat h^s + \hat h^d + \hat h^g +
\hat h^{sg} + \hat h^{gd}$ where:

\bea
 \hat h^s & = & \sum_{l = 1}^\infty \left[\e_s \hat s_l^\dagger \hat s_l +  t_s \left(
  \hat s_l^\dagger  \hat s_{l+1} + {\rm h.c.}\right)\right],\nn\\
   \hat h^g & = & \sum_{l = 1}^n \left[ \e_g \hat g_l^\dagger \hat g_l   +  t_g \left(
    \hat g_l^\dagger  \hat g_{l+1} + {\rm h.c.}\right)\right],\nn\\
     \hat h^{sg} & = & t_{sg} \left(\hat s_1^\dagger  \hat g_1 + {\rm h.c.}\right),
      \lb{eq9}
       \eea

\begin{figure}
\center\includegraphics[width=9cm]{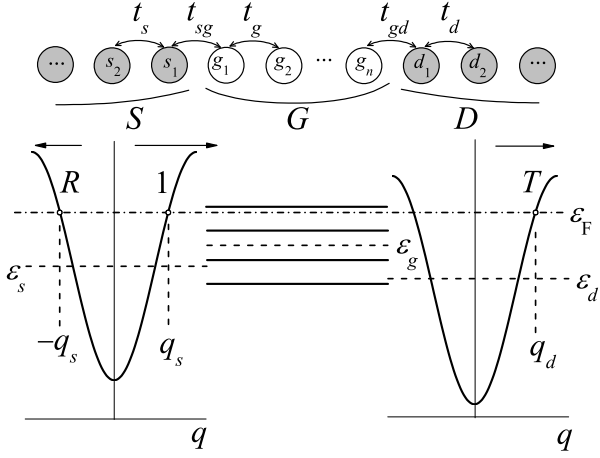}
\caption{Composite system of finite $n$-chain (gate element, $G$) inserted between two
semi-infinite chain leads (source, $S$, and drain, $D$). The energy diagram shows the
on-site energy levels (dashed) for $i$-th element ($i=s, g,d$) and the Fermi level
(dot-dashed) whose crossings with the continuous S- and D- dispersion curves define the
wave numbers for incoming ($q_s$), reflected ($-q_s$) and transmitted ($q_d$) parts of the
Fermi state. Notice that the Fermi level generally does not match any of the discrete levels
(solid) in the central (G) element.}
\lb{cap:2}
\end{figure}

\noindent including the respective on-site energies $\e_i$ ($i = s,d,g$) and hopping parameters
$t_i$ ($i = s,sg,g,gd,d$), while the operators $\hat h^d$, $\hat h^{gd}$ are analogous to
$\hat h^s$, $\hat h^{sg}$ with the formal change of indices $s \to d$. For this macroscopic
system, the energy spectrum includes continuous S- and D-bands $\e_{i,q} = \e_i + 2 t_i \cos q,
\,i=s, d$ and possibly discrete G-levels outside these bands. The collective eigen-state for a
given energy $\e$ can be found from the equations of motion that generalize Eq. \ref{eq6}. We
denote $s_l,\, g_l$ or $d_l$ the respective local amplitudes of the wave function and define
the dimensionless dynamical variables $x_i = \left(\e - \e_i \right)/2t_i$ ($i=s,g,d$). Let
the S-amplitude be a sum of an incident wave of intensity 1 with the wave number $q_s = \arccos
x_s$ and a reflected wave with certain amplitude $R$ and the wave number $-q_s$:

\be
 s_l = {\rm e}^{-i q_s l} + R {\rm e}^{i q_s l}
  \lb{eq10}
   \ee

\noindent (for regressive order of sites $l$ in S), and the D-amplitude present a transmitted
wave with certain amplitude $T$ and the wave number $q_d = \arccos x_d$:

\be
 d_l = T{\rm e}^{i q_d l}.
  \lb{eq11}
   \ee

\noindent Eqs. \ref{eq10} and \ref{eq11} refer to one of fundamental solutions for given
$\e$ (besides that where the incident and reflected waves belong to D and the transmitted
one does to S). These forms automatically satisfy the equations of motion within S and D:

\be
 2x_s s_l  =  s_{l-1} + s_{l+1},\quad  2 x_d d_l = d_{l-1} + d_{l+1}
  \lb{eq12}
    \ee

\noindent (for $l \geq 2$), while the pairs of equations on the S/G and G/D interfaces:

\bea
  2s_1 \cos q_s & = & s_2 + \frac{t_{sg}}{t_s}g_1,\nn\\
   u_1 g_1 & = & g_2 +  \frac{t_{sg}}{t_g} s_1,
    \lb{eq13}
    \eea

\noindent and

 \bea
  2d_1 \cos q_d & = & d_2 + \frac{t_{gd}}{t_d}g_n,\nn\\
 u_1 g_n & = & g_{n-1} + \frac{t_{gd}}{t_g} d_1,
    \lb{eq14}
      \eea

\noindent are the discrete analogs of usual boundary conditions for continuous wave function
and its derivative \cite{slonczewski}. They permit to express the terminal pairs of
G-amplitudes through the asymptotic parameters $R,\,T,\,q_s$ and $q_d$:

\bea
 g_1 & = & \frac{t_s}{t_{sg}} \left(1 + R\right),  \quad  g_2  =  \frac{t_s}{t_{sg}}
 \left[u_1 - \gamma_s^\ast +  \left(u_1 - \g_s \right)R \right],\nn\\
  g_n & = & \frac{t_d}{t_{gd}}T, \qquad \qquad g_{n-1} =  \frac{t_d}{t_{gd}}\left(u_1 -
   \g_d \right)T,
    \lb{eq15}
      \eea

\noindent with the interface parameters $\g_s = {\rm e}^{i q_s} {t_{sg}}^2/t_g t_s$ and
$\g_d = {\rm e}^{i q_d}{t_{gd}}^2/t_g t_d$. The polynomials $u_l \equiv u_l(x_g)$ are
formally the same as given by Eq. \ref{eq4} with the energy argument $x_g = (\e - \e_g)/2
t_g $. But the energies $\e$ of our main interest for the transport processes are those close
to the Fermi energy $\e_{\rm F}$ which is generally \emph{not} an eigenvalue, Eq. \ref{eq5},
for the isolated G-element. Therefore the transient "momentum" $q_g = \arccos x_g$ (not
necessarily real) breaks down the closed boundary conditions, Eq. \ref{eq8}, for G and thus
enables continuity of quantum states along the composite system. Next, using Eq. \ref{eq6}
for this element in the form:

\be
 u_1 g_l = g_{l+1} + g_{l-1},
  \lb{eq16}
   \ee

\noindent it is possible to interrelate the terminal G-amplitudes:

\bea
  g_{n-1} & =  &  u_{n-2} g_1 - \frac{t_{sg}}{t_g} u_{n-3}s_1, \nn\\
   g_{n}  & = &   u_{n-1} g_1 - \frac{t_{sg}}{t_g} u_{n-2}s_1.
  \lb{eq17}
    \eea

\noindent Then, Eqs. \ref{eq15} and \ref{eq17} yield two independent relations between
the coefficients $R$ and $T$. Those are readily solved to give:

\bea
 R\left(x_s,x_g,x_d\right)  & = & - \frac{\overline D_n}{D_n}, \nn\\
  T\left(x_s,x_g,x_d\right) & = & - \frac{2i \sqrt{\left|\g_s\g_d/t_st_d\right|}}{D_n},
   \lb{eq18}
     \eea

\noindent where the resonance properties result from the denominator:

\be
 D_n\left(x_s,x_g,x_d\right) = u_n -  \left(\g_s +\g_g\right)u_{n-1} + \g_s \g_d u_{n-2},
  \lb{eq19}
   \ee

\noindent with the relevant variables $x_i$ as arguments of complex factors $\g_i$ and
real polynomials $u_l$, and $\overline D_n(x_s,x_g,x_d) \equiv D_n(x_s + \pi,x_g,x_d)$.
Since, in the considered 1D case, all $x_i(\e) = \left(\e - \e_i\right)/2t_i$ are defined
by the single energy variable $\e$, the coefficients $R$ and $T$ can be also defined as
functions of energy: $R(\e) \equiv R\left(x_s(\e),x_g(\e),x_d(\e)\right)$ and $T(\e)\equiv T\left(x_s(\e),x_g(\e), x_d(\e)\right)$. It is important to notice that the result of Eqs.
\ref{eq18},\ref{eq19} is just analogous to that obtained with the Green function techniques
\cite{mathon1}, the factors $\g_s$ and $\g_d$ playing the role of interface Green functions
of Ref. \cite{mathon1}. A typical behavior of the transmission coefficient $|T(\e)|^2$ is
presented in Fig. \ref{cap:3}. It shows $n$ transmission resonances generated by $n$
discrete energy levels of the G-element (by $n$ atoms in the chain) as they are passing
over the Fermi level within the mutually displaced energy bands. The displacement can be due,
for instance, to the Stoner splitting between majority and minority subbands of oppositely
polarized S- and D-elements (see also Sec. \ref{3D}). Notice that the resonances become
sharper as the levels approach the band edges, and the maximum transmission in the asymmetric
S-D band configuration is not limited to unity. This coefficient enters the Landauer formula
\cite{landauer} for the ballistic conductance through the 1D composite system (in zero
temperature limit):

\begin{figure}
\center\includegraphics[width =9.5 cm]{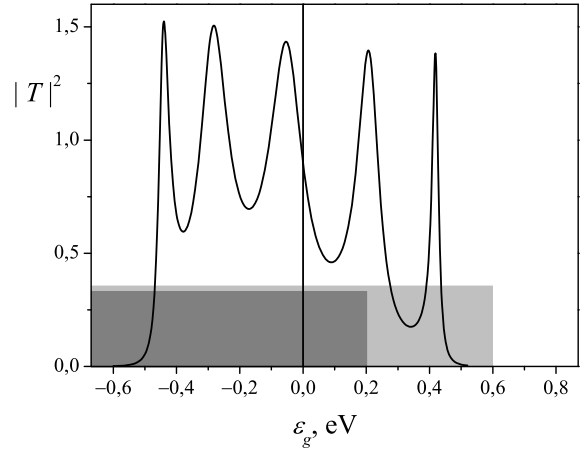}
\caption{Transmission coefficient $|T|^2$ in function of the on-site energy $\e_g$ in the
gate element of the composite chain system, for the choice of its parameters $\e_s = -
0.4$ eV, $\e_d = -0.8$ eV (relative to the Fermi energy), $t_s = t_d = 0.5$ eV, $t_g =
t_{sg} = t_{gd} = 0.25$ eV and $n = 5$. The shadowed areas indicate the (Stoner shifted)
continuous bands, S (light grey) and D (dark grey).}
 \lb{cap:3}
\end{figure}

\be
 G = \frac{e^2}{h}|T\left(\e_{\rm F}\right) |^2,
  \lb{eq20}
   \ee

\noindent with the Fermi energy $\e_{\rm F}$. Now, allowing the S and D chains to support
spin polarized subbands $\e_{i,q,\s} = \e_{i,\s} + 2 t_i \cos q$ (where $\e_{i,\s} = \e_i -
\s \D_i$, $\s = \pm$ are the majority and minority spin indices and $\D_i$ are the Stoner
splitting parameters for $i = s,d$), we can introduce the energy and spin-dependent variables
$x_{i,\s}(\e) = \left(\e - \e_{i,\s}\right)/\left(2t_i\right)$, $i = s,d$, for in- and
out-channels and obtain from Eq. \ref{eq20} the spin-dependent conductance values
$G_{\s,\s^\prime} = \left(e^2/h\right) |T\left(x_{s,\s}\left(\e_{\rm F}\right),x_{d,
\s^\prime}\left(\e_{\rm F}\right), x_g\left(\e_{\rm F}\right)\right)|^2$. Finally, the
(maximum) magnetoresistance is defined as usually through the difference between the
conductance values $G_{P} = G_{+,+} + G_{-,-}$ for parallel and $G_{AP} = G_{+,-} + G_{+,-}$
for antiparallel S/D polarization:

\be
 MR = \frac{G_{P}}{G_{AP}} - 1.
  \lb{eq21}
   \ee

\noindent Although the state-of-the-art technology already permits development of such
genuinely 1D devices \cite{agrait} and the resonance behavior like that in Fig. \ref{cap:3}
(different from the known quantized conductance steps vs voltage bias) can be directly sought
in them, it is of major practical importance to generalize the above treatment for a more
realistic multilayered structure and this will be done in the next section.

\section{3. 3-dimensional multilayered structure}

\lb{3D} Passing from 1D composite chain to multilayered (and spin polarized) 3D lattice structure
as shown in Fig. \ref{cap:4}, we extend the indexing of site operators from $\hat s_l,\,
\hat d_l$ and $\hat g_l$ to $\hat s_{l,\bm,\s},\,\hat d_{l,\bm,\s}$ and $\hat g_{l,\bm,\s}$,
where $\bm$ runs over $N$ sites in the $l$th plane and $\s$ is $\pm$. Our strategy in this
case relies on the conservation of the transversal quasi-momentum $\bk = \left(k_x, k_y\right)$
in the transitions between the planes \cite{mathon, itoh2}. From the experimental point of
view, this requires perfect interfaces that are only reachable with advanced molecular beam
epitaxy (MBE) techniques \cite{yuasa3}. To describe the situation where $\bk$ is a good quantum
number for independent 1D-like longitudinal transport channels, we pass from the site operators
to the "planar wave" operators. Thus, for the $l$th plane in the S element, the latter operators
are defined as:

\be
  \hat s_{l,\bk,\s} = \frac 1 {\sqrt N}  \sum_\bm  {\rm e}^{i \bk \cdot \bm} \hat s_{l,\bm,\s},
    \lb{eq22}
      \ee

\begin{figure}
\center\includegraphics[width=9cm]{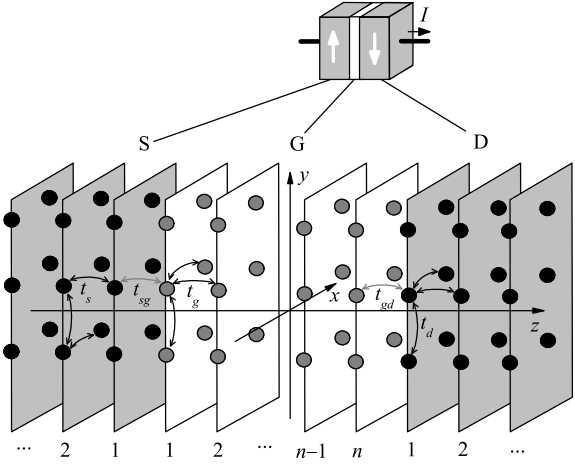}
\caption{Real multilayered structure where the current $I$ flows through two ferromagnetic
electrodes, S and D, separated by a non-magnetic spacer G and its model by the composite 3D
system where a finite $n$-plane spacer is inserted between two semi-infinite leads.}
 \lb{cap:4}
\end{figure}

\noindent and, similarly, the "planar wave" operators  $\hat d_{l,\bk,\s}$ and $\hat g_{l,
\bk,\s}$ for D and G elements are written. The related extension of the Hamiltonian is:
$\hat H = \sum_{\bk,\s}\left(\hat h_{\bk,\s}^s + \hat h_{\bk,\s}^d + \hat h_{\bk,\s}^g +
\hat h_{\bk,\s}^{sg} + \hat h_{\bk,\s}^{gd}\right)$, where the particular terms are analogous
to those in Eq. \ref{eq9} with the change of all the site operators by the "planar wave" ones
and all the on-site energies $\e_i$ by the transversal momentum subbands $\e_{i,\bk,\s} =
\e_{i,\s}  + 2t_i\left(\cos k_x + \cos k_y\right), i = s,d$ and $\e_{g,\bk,\s} = \e_g + 2t_g
\left(\cos k_x + \cos k_y\right)$. The equations of longitudinal motion in terms of the
"planar wave" amplitudes $s_{l,\bk,\s}$, $d_{l,\bk,\s}$ and $g_{l,\bk,\s}$ (for given energy
$\e$ of the collective state) are obtained in analogy with the 1D case. Thus, in the leads S
and D (beyond the interfaces, at $l > 1$), they are analogs to Eq. \ref{eq12}:

\bea
 2x_{s,\bk,\s} s_{l,\bk,\s} & = & s_{l-1,\bk,\s} + s_{l+1,\bk,\s},\nn\\
  2x_{d,\bk} d_{l,\bk,\s} & = & d_{l-1,\bk,\s} + d_{l+1,\bk,\s},
   \lb{eq23}
    \eea

\noindent with $x_{i,\bk,\s} = \left(\e - \e_{i,\bk,\s}\right)/\left(2t_i\right)$, $i = s,d$,
while in the spacer G (at $1 < l < n$), we have in analogy with Eqs. \ref{eq7} and \ref{eq15}:

\be
 2x_{g,\bk} g_{l,\bk,\s} = g_{l-1,\bk,\s} + g_{l+1,\bk,\s},
  \lb{eq24}
    \ee

\noindent with $x_{g,\bk} = \left(\e - \e_{g,\bk}\right)/\left(2t_g\right)$. Also the
equations for interface amplitudes:

\bea
 2 x_{s,\bk,\s} s_{\bk,1,\s} & = & s_{\bk,2,\s} + \frac{t_{sg}}{t_s} g_{\bk,1,\s},\nn\\
  2 x_{g,\bk,\s} g_{\bk,1,\s} & = & g_{\bk,2,\s} + \frac{t_{sg}}{t_g} s_{\bk,1,\s},\nn\\
   2 x_{g,\bk,\s} g_{\bk,n,\s} & = & g_{\bk,n-1,\s} + \frac{t_{gd}}{t_g} d_{\bk,1,\s},\nn\\
    2 x_{d,\bk,\s} d_{\bk,1,\s} & = & d_{\bk,2,\s} + \frac{t_{gd}}{t_d} g_{\bk,1,\s},
     \lb{eq25}
      \eea

\noindent are analogous to Eqs. \ref{eq13}, \ref{eq14} and \ref{eq15}. The next derivation,
in full similarity with the 1D case, leads to the full dispersion laws in the leads
$\e_{i,\bk,q,\s} = \e_{i,\bk,\s} + 2t_i\cos q$ (for $i = s,d$) and to the final conductance
formula for particular in-out spin channels:

\be
 G_{\s \s'} = \frac{e^2}{h}\sum_{\bk\in K}|T_{\s \s'}(\e_{\rm F},\bk)|^2.
 \lb{eq26}
  \ee

\noindent Here the transmission coefficient depends on the relevant variables $\s, \s', \e$
and $\bk$ accordingly to: $T_{\s \s'}(\e,\bk) \equiv T(q_{s,\bk,\s}, q_{g,\bk}, q_{d,\bk,
\s^\prime})$ with $q_{i,\bk,\s} = \arccos x_{i,\bk,\s}$ for $i = s,d$ and $q_{g,\bk} =
\arccos x_{g,\bk}$, and the sum in $\bk$ is restricted to the "permitted" range $K$, such
that simultaneous equalities $\e_{s,\bk,q_s} = \e_{d,\bk,q_d} = \e_{\rm F}$ are possible for
certain \emph{real} $q_s$ and $q_d$. In more detail, the latter condition is expressed as:

\begin{widetext}
\be
 \max\left\{-2,\max\left[x_{s,\s}\left(\e_{\rm F}\right),x_{d,\s^\prime}\left(\e_{\rm F}\right)
  \right]-1 \right\} \leq \cos k_x + \cos k_y  \leq \min \left\{2,\min\left[x_{s,\s}\left(
   \e_{\rm F}\right),x_{d,\s^\prime}\left(\e_{\rm F}\right)\right] + 1\right\},
    \lb{eq27}
     \ee
\end{widetext}

\noindent fully defining the integration procedure (in the limit of continuous $\bk$). Then,
seeking for optimum performance of the model MR device from Eq. \ref{eq21}, it is of interest
to evaluate it as a function of the system parameters, mainly the number of atomic layers in
the gate  $n$ and the on-site energy level of the gate $\e_g$ (which can be possibly controlled
through the gate bias). Also, variation of the latter parameter from positive to negative values
permits to model in our approach the passage from the tunnel magnetoresistance (TMR) to giant
magnetoresistance (GMR) regime in a unified way.

The following numerical work can be oriented accordingly to some evident qualitative arguments.
The variation of the integrand in Eq. \ref{eq26} is mainly controlled by that of the polynomials
$u_l\left(x_{g,\bk}\right)$ in the denominator of Eq. \ref{eq19}. As seen from the explicit
Eq. \ref{eq4}, they are oscillating if $\left|x_{g,\bk}\right| < 1$ (that is, the sampling
point $\e_{g,\bk}$ in the G-spectrum close enough to the Fermi energy $\e_{\rm F}$) and
exponentially growing if $\left|x_g\right| > 1$ (remote enough $\e_{g,\bk}$ from $\e_{\rm F}$).
Therefore, the conductance is generally expected to oscillate (either in $\e_g$ and in $n$) as
far as the level $\e_g$ is close enough to $\e_{\rm F}$ (which can be compared to the GMR regime)
and to exponentially decay at $\e_g$ far enough from $\e_{\rm F}$ (a generalized TMR regime). The
latter decay should asymptotically tend to $MR(n) \propto \exp\left(-n x_{\min}\right)$ with
$x_{\min} = \min_{\bk \in K}\left|x_{g,\bk}\right|$ at $n \gg 1$.

In the latter case, the direct calculation by Eqs. \ref{eq21} may result in $G_P$ and $G_{AP}$
both exponentially small but the latter yet much smaller and thus in (arbitrarily) huge MR
values. However, they should not be physically attainable, taking into account that the real
multiband electronic structure of transition metals always includes some additional conduction
channels, for instance due to the \emph{s}-bands, whose tunnel contribution slower decays than
that of \emph{d}-bands and is almost spin independent. Therefore it should dominate the
transport in the indicated regime and make the real MR exponentially small. A simple
phenomenological account of this mechanism in the considered single-band model can be done by
introducing a certain spin-independent term $G_0$ into either $G_P$ and $G_{AP}$ values:

\be
 G_{P} = G_{++} + G_{--} + G_0, \quad G_{AP} = G_{+-} + G_{-+} + G_0,
 \lb{eq28}
  \ee

\noindent to present the MR formula, Eq. \ref{eq21} as

\be
  MR = \frac{G_{++} + G_{--} - G_{+-} - G_{-+}}{G_{+-} + G_{-+} + G_0}.
  \lb{eq29}
   \ee

\begin{figure}
\center\includegraphics[width =9.5 cm]{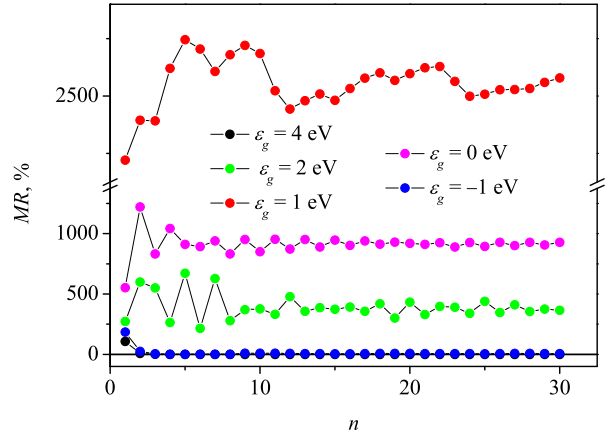}
\caption{Magnetoresistance of a FM/NM/FM junction with the model parameters: $\e_{s,+} =
\e_{d,+} = 1.32$ eV, $\e_{s,-} = \e_{d,-} = 3.36$ eV, $t_s = t_d = t_g = -0.6$ eV  (like
those from Ref. \cite{mits}) and $\gamma_{s,d} = 0.5$ in function of the number $n$ of
spacer layers at fixed values of $\e_{g}$. Notice the exponential decay in the TMR-like regime
either at the highest $\e_{g} = 4$ eV and the lowest $\e_{g} = -1$ eV and a strong enhancement
with emergence of oscillatory behavior at intermediate $\e_{g}$ ("shallow band" regime).}
 \lb{cap:5}
\end{figure}

\begin{figure}
\center\includegraphics[width =9.5 cm]{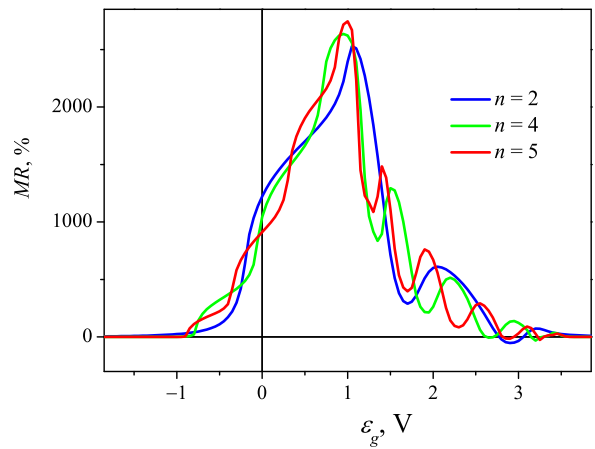}
\caption{MR vs the on-site energy $\e_g$ of the gate element for a junction with the same
parameters as in Fig. \ref{cap:5} and various numbers of atomic planes in the gate element,
$n = 2$, $n = 4$ and $n = 5$. Compare the resonance peaks in the "shallow band" regime with
those in the 1D case of Fig. \ref{cap:3}.}
\lb{cap:6}
\end{figure}

\noindent It is just this formula that is used below for all practical MR calculations. Thus,
using the band structure parameters: $\e_{s,+} = \e_{d,+} = 1.32$ eV, $\e_{s,-} = \e_{d,-} =
3.36$ eV, $t_s = t_d = t_g = -0.6$ eV and $\gamma_{s,d} = 0.5$ (a reasonable single-band fit
for the real Fe band structure, see \cite{callaway,nautial,mits}) and choosing for simplicity
the constant value $G_0 = 0.1e^2/h$, we find that the MR behavior vs $n$ indeed changes
qualitatively at different choices of $\e_g$ (Fig. \ref{cap:5}). The TMR-like behavior with
fast exponential decay appears either at high enough gate level, $\e_g \gtrsim 6t_g$ (which can
be compared to a "positive" barrier in the continuum approximation), and at low enough $\e_g
\lesssim -2t_g$ (a "negative" or "hole" barrier), but it develops GMR-like oscillations with
greatly increasing overall MR amplitude at the intermediate $\e_g$ values (which can be called
the "shallow band" regime). The oscillating behavior is in a qualitative agreement with that
experimentally observed for MgO moderate tunnel barriers between Fe electrodes \cite{yuasa1},
except for stronger first oscillations than in the data. However, it will be shown below that
these strong oscillations are effectively moderated when taking into account the specific
interface effects between metal and insulator layers. The most notable feature of the
calculated MR is its amazingly high maximum value, of the order of 3000 \%, indicating a
huge potentiality of the quantum coherent conduction regime.

For the same choice of parameters, the calculated dependencies of MR vs $\e_g$ (at fixed
values of $n = 2,4,5$) are shown in Fig.\ref{cap:6}. In concordance with the above
considered $MR(n)$ behavior, they practically vanish beyond the range of intermediate $\e_g$
and display a finite number of resonance peaks within this range (reminiscent of $n$ 1D
resonances in Fig. \ref{cap:3}), reaching the same highest order of magnitude in the "shallow
band" regime. These very high values in the present tight-binding approach contrast with the
known result for the model of almost free electrons on the continuous rectangular barrier
\cite{slonczewski}, where MR reaches zero minimum at low barrier height. As yet, the
$MR\left(\e_g\right)$ dependence was only studied experimentally for Al-O spacers
\cite{tezuka}, possessing most probably polycrystalline or amorphous structure and high enough
$\e_g$, so it could be of interest to try it also with epitaxial MgO spacers and possibly with
those spacer materials that can realize the "shallow band" condition.

\section{4. Interfacial Effects}

\lb{delta} In this section, we will discuss the interfacial effects present at the metal/insulator
or metal/non-magnetic-spacer interfaces. This is motivated by the analogy with the well known
case of Schottky barrier at metal/semiconductor interfaces, leading to such interesting
physical effects as band bending \cite{sze}. It is known from X-ray and ultraviolet
photoemission spectroscopy (XPS and UPS) that some charge transfer effects also appear at
the metal-insulator interface, leading to formation of an interfacial charge-dipole whose
magnitude is defined by the localized states at interfaces \cite{popinciuc}. Since this
dipole directly affects the efficiency of tunneling, it is also important to evaluate its
effect in the magnetoresistance.

The best treatment of this problem is to introduce self-consistently a charging energy
($\d$, commonly called the built-up voltage) due to a charge accumulation in the framework
of the density functional theory (DFT). This is going to be done in the future work, but at
the moment we will develop simple analytic formulas to take into account these interfacial
effects qualitatively. Despite of its simplicity, the model can exemplify in which way the
formation of charge dipoles affects the magnetoresistance ratio.

\begin{figure}
\center\includegraphics[width =8 cm]{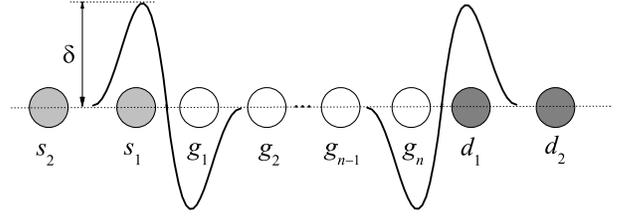}
\caption{Schematic representation of the interface charge-energy, $\d$, created by a charge
accumulation in the S/G and G/D interfaces, as a simplified description of the true
self-consistent behavior.}
 \lb{cap:7}
\end{figure}

We go on using the same model of Sec. \ref{3D} but considering extra charge energies $\pm
\d$ on the sites pertaining to the two atomic planes on both sides of each interface (see
Fig. \ref{cap:7}) as an approximation for more realistic charge and potential distributions
around interfaces, obtained by numerical DFT calculations \cite{butler}. The $\d$-perturbation
results in new boundary conditions and, as a consequence, in a new transmission coefficient.
In this approximation, there is no changes in equations of motion within the particular
elements (S, D and G), but new pairs of equations appear at the S/G and G/D interfaces,
involving the charge energy $\d$:

\bea
 \left(2\cos q_s+ \d/t_s\right) s_1 &=& s_2 + \left(t_{sg}/t_s\right)g_1,\nn\\
  \left(x_g - \d/t_g\right)g_1 & = & g_2 + \left(t_{sg}/t_g\right)s_1,\nn\\
   \left(2\cos q_b + \d/t_d\right) d_1 & = & d_2 + \left(t_{gd}/t_d\right)g_n,\nn\\
    \left(x_g- \d/t_g\right) g_n & = & g_{n-1} + \left(t_{gd}/t_g\right)d_1.
     \lb{eq30}
      \eea

\begin{figure}
\center\includegraphics[width =9 cm]{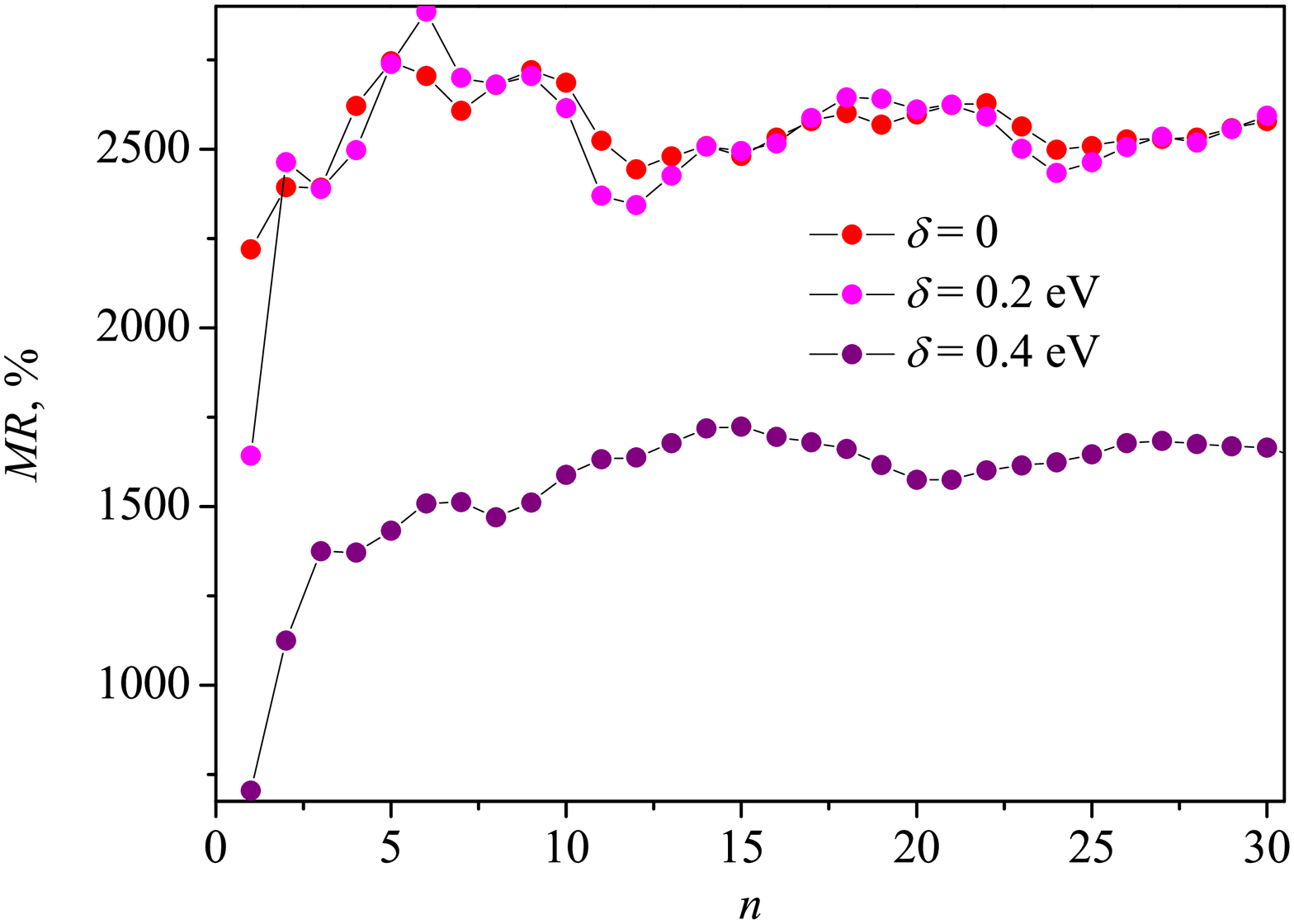}
\caption{Effect of the interface charge energy on the $MR(n)$ dependence. The system
corresponds to the parameters as in Figs. \ref{cap:5}, \ref{cap:6} with $\e_g = 1$ eV and
$\d$ varying from $0$ to $0.4$ eV. }
\lb{cap:8}
\center\includegraphics[width =9 cm]{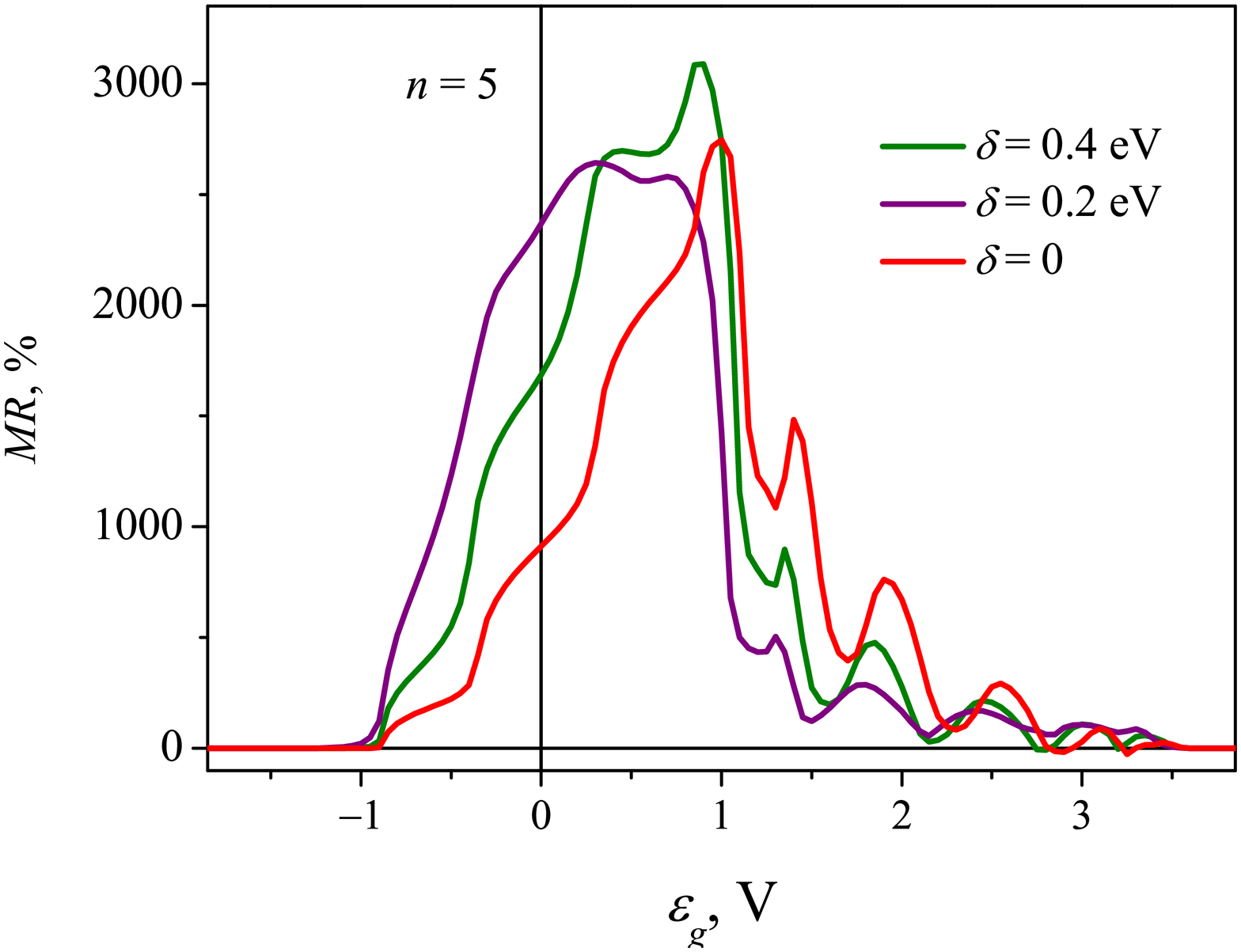}
\caption{Effect of the interface charge energy on the $MR(\e_g)$ dependence. The system
corresponds to the parameters as in Figs. \ref{cap:5}, \ref{cap:6} with $n = 5$ and $\d$
varying from $0$ to $0.4$ eV. }
\lb{cap:9}
\end{figure}

\noindent These boundary conditions allow to re-calculate two terminal G-amplitudes in
function of the parameters $R,\,T,\,q_s, q_d$. Interconnecting these terminal amplitudes
leads to the transmission formula, Eq. \ref{eq18}, but with the modified denominator $D_{n,\d}
= A_n - B_n + C_n$ where:

\bea
 A_n & = & \left( 1+ \frac{\d}{t_d}e^{iq_d}\right)\left(1+\frac{\d}{t_s}e^{iq_s}\right)\nn\\
  &\times&\left(u_n-2\frac{\d}{t_g}u_{n-1}+\left(\frac{\d}{t_g}\right)^2u_{n-2}\right), \nn\\
   B_n & = & \left(\g_s\left( 1+ \frac{\d}{t_d}e^{iq_d}\right) + \g_d\left(1+\frac{\d}{t_s}
    e^{iq_s}\right) \right)\nn\\
     &\times&\left(u_{n-1}-\frac{\d}{t_g}u_{n-2}\right), \nn\\
      C_n & = &\g_s\g_d u_{n-2}.
       \lb{eq31}
        \eea

\noindent It is easy to see that $D_{n,\d} \to D_n$ in the limit of $\d \to 0$.

The MR defined from Eqs. \ref{eq29},\ref{eq31} in function of the number $n$ of gate atomic
planes and of the gate voltage $\e_g$ for tree values of the interface potential  $\d$ are
presented in Figs. \ref{cap:8}, \ref{cap:9}. The obtained softening of first oscillations
makes these curves more similar to the experimental observations \cite{yuasa1}. An unexpected
result is that the effect of an extra barrier due to the charge energy can yet reinforce the
calculated MR peak in the shallow band regime, though reducing the values in the TMR regime
at higher barrier height. Obviously, the charge energy barrier reduces the conductance (either
in P and AP configurations), but the MR enhancement is mainly due to a much stronger reduction
of the AP conductance. Apparently it results from the wave function localization caused by
coherent resonances in the interfacial potential wells. This idea of charge energy induced
resonances is corroborated by the calculated sharpening of peaks just in the AP conductance.
Amazingly high peak MR values, reaching $\sim 3000 \%$ for a reasonable choice of $\d \sim 0.4$
eV (similar to the numerical estimate for Fe-MgO interfaces, \cite{butler}), should motivate
fabrication of new devices with the choice of such spacer materials as semiconducting (Ge, Si)
or semimetalic (Sb, As). Though the peak value may be obviously decreased under the effects of
electron-electron, electron-phonon, and electron-magnon interactions, finite temperature and
disorder, it expresses one of the principal results of this work, demonstrating that the highest
possible MR value should be reached in the shallow-band regime for non-magnetic spacer by adjusting
to the strongest resonance condition, once electronic coherence is assured.

\section{5. Conclusions}

\lb{conc} In this work a theoretical approach was developed to fully coherent spin-dependent quantum
transport in nanolayered magnetic junctions, using single-band tight-binding model with
explicit equations of motion for wave-function amplitudes. The analytic solutions for the
transmission and reflection coefficients were generalized for a 3-dimensional magnetic
junction structure. The simple zero temperature calculations have revealed the most
pronounced enhancement of the magnetoresistance in the "shallow band" regime, defined by
low gate voltages (Fig. \ref{cap:6}). Another important feature for this gate voltage regime
is the calculated oscillatory behavior of MR with the number of atomic planes in the spacer.
In support of these theoretical predictions, the calculations also reveal that the oscillatory
regime starts already at moderate gate voltages, $\e_g \sim 2$ eV. This agrees rather well
with the experimental observation by Yuasa \cite{yuasa1} of clear MR oscillations at low
enough gate voltage barrier $\e_g \sim 0.4$ eV in a Co/MgO/Co structure. So it is concluded
that the best MR values for a quantum magnetic junction could be reached using shallow band
materials for spacer layers, the possible candidates sought between transition metals (Cr
\cite{greullet} in junctions of the type Fe/Cr/Fe or Zn in junctions of the type Co/Zn/Co),
semiconductors (Ge, Si), or semimetals (Sb, As). Finally, the important effect of charge
build-up (Sec. \ref{delta}) on the junction interface was also considered and, though in a
simple phenomenological approach, a possibility is shown for even stronger enhancement of
the magnetoresistance in presence of the extra charge-energy for the same shallow band regime,
emphasizing again the promise of using the low $\e_g$-materials. To verify these model
predictions, the future work must include various realistic effects, such as those of finite
temperature and self-consistent density functional theory (DFT), to adopt also the multiband
electronic structure, spin-transfer processes, and non-linear conduction.

\section{Acknowledgements}

The authors wish to thank J.B. Sousa, P.P. Freitas, J. Lopes dos Santos, J. P. Araújo and
H. L. Gomes for their helpful discussions and friendly support of various aspects of this
study. The work was partially supported by Portuguese Foundation of Science and Technology,
FCT, through the grant SFRH/BD/24190/2005 (H. S.).

\end{document}